\documentclass{article} %
\usepackage[final]{colm2026_conference}

\usepackage{microtype}
\usepackage{hyperref}
\usepackage{url}
\usepackage{booktabs}
\usepackage{amssymb}

\usepackage{lineno}

\definecolor{darkblue}{rgb}{0, 0, 0.5}
\hypersetup{colorlinks=true, citecolor=darkblue, linkcolor=darkblue, urlcolor=darkblue}

\usepackage{hyperref}
\usepackage{url}

\usepackage{xspace}
\usepackage{lipsum}
\usepackage{multirow}
\usepackage{multicol}
\usepackage{colortbl}
\usepackage{listings}
\usepackage{enumitem}
\usepackage{tabularx}
\usepackage{booktabs}
\usepackage{array}
\usepackage{caption}
\usepackage{enumitem}
\usepackage{caption}
\usepackage{circledsteps}
\usepackage{vcell}
\usepackage{bbm}
\usepackage{wrapfig}
\usepackage{amsmath}
\usepackage{cleveref}
\usepackage[many]{tcolorbox}
\usepackage{booktabs}
\usepackage{circledsteps}
\usepackage[T1]{fontenc}

\newcommand{\name}[0]{CollabSkill\xspace}

\crefformat{section}{\S#2#1#3} %
\crefformat{subsection}{\S#2#1#3}
\crefformat{subsubsection}{\S#2#1#3}

\newcommand{\reffig}[1]{Figure~\ref{#1}}
\newcommand{\refsec}[1]{\S\ref{#1}} %
\newcommand{\reftab}[1]{Table~\ref{#1}}

\def\eg{\textit{e.g.}\xspace}

\def\ie{\textit{i.e.}\xspace}

\newcommand{\RNum}[1]{\uppercase\expandafter{\romannumeral #1\relax}}

\definecolor{myyellow}{RGB}{249, 231, 173}
\definecolor{mygreen}{RGB}{233, 242, 230}
\definecolor{mygrey}{RGB}{250, 250, 250}
\definecolor{myblue}{RGB}{245, 248, 250}%
\definecolor{communication}{RGB}{224, 242, 254}
\definecolor{situation}{RGB}{220, 252, 231}
\definecolor{planning}{RGB}{255, 237, 213}
\definecolor{environment}{RGB}{243, 232, 255}
\definecolor{personalization}{RGB}{254, 226, 226}

\newtcolorbox{promptbox}[1]{
    enhanced,
    breakable,
    boxrule=1pt,  %
    fontupper=\small,
    fonttitle=\bfseries\color{white},
    arc=3pt,  %
    rounded corners,
    colframe=black,
    colbacktitle=gray!50!black,
    colback=mygrey,
    title=#1,
    left=2mm,  %
    right=2mm,  %
    top=1mm,  %
    bottom=1mm  %
}

\newtcolorbox{examplebox}[1]{
    enhanced,
    breakable,
    boxrule=1pt,  %
    fontupper=\small,
    fonttitle=\bfseries,
    arc=0mm,  %
    rounded corners,
    colframe=black,
    colbacktitle=gray!50!black,
    colback=yellow!10,
    title=#1,
    left=2mm,  %
    right=2mm,  %
    top=1mm,  %
    bottom=1mm,  %
    breakable
}

\newtcolorbox{agentbox}[1]{
    enhanced,
    breakable,
    boxrule=1pt,  %
    fontupper=\small,
    fonttitle=\bfseries\color{white},
    arc=3pt,  %
    rounded corners,
    colframe=black,
    colbacktitle=gray!50!black,
    colback=myblue,
    title=#1,
    left=2mm,  %
    right=2mm,  %
    top=1mm,  %
    bottom=1mm  %
}

\title{\name: Evaluating Human-Agent Collaboration\\ On Real-World Tasks}

\author{
  Yijia Shao$^1$ \quad
  Zora Z. Wang$^2$ \quad
  Neel Ahuja$^1$ \quad
  Yicheng Wang$^3$ \quad
  Bowen Liu$^4$ \\[0.5em]
  ~\textbf{Diyi Yang}$^1$ \\[0.5em]
  $^1$Stanford University \quad
  $^2$Carnegie Mellon University\\[0.5em]
  $^3$University of Rochester \quad
  $^4$Individual Researcher \\[0.5em]
  {\texttt{\{shaoyj, diyiy\}@cs.stanford.edu}} \\
  [0.5em]Website: \url{https://cogym.saltlab.stanford.edu}
}

\begin{document}

\ifcolmsubmission
\linenumbers
\fi

\maketitle

\begin{abstract}
AI agents are reshaping the workspace, leading to drastic change of how humans work. Despite the considerable potential of human-agent collaboration both in preserving human agency and generating economic value, this paradigm remains largely absent from occupational task evaluation, hindered by the difficulty of gathering real human data and accounting for inter-human variability. We introduce \textbf{\name}, a framework for evaluating human-agent collaboration on real-world occupational tasks. \name pairs real human workers with AI agents on tasks matched to their occupational background, collecting data that capture the complexity of economically valuable tasks and the usage patterns of real workers. To account for inter-human variability, \name employs a Bayesian skill rating system to disentangle and quantify the skill contributions of both humans and AI agents. Drawing on over 1,500 prompts from 386 working sessions contributed by 93 human workers, our analysis yields insights on two fronts: on the agent side, rankings on \name diverge from those of existing fully autonomous benchmarks where Codex leads, with Claude Code ranking first; on the human side, \name reveals that practical experience emerges as the primary driver of collaboration skill, with hands-on collaboration meaningfully shifting workers' AI literacy. Together, we hope \name enables the community to invest in systematic evaluation of human-agent collaboration and spurs development efforts aimed at building AI agents that genuinely augment human workers.

\end{abstract}

\section{Introduction}
AI agents are bringing unprecedented opportunities but also massive disruptions to the workspace, raising pressing concerns around job loss, inequality, and reduced human agency~\citep{handa2025economic,demirci2025ai,hazra2025position,hoffmann2025generative}. Human-agent collaboration represents a promising yet underexplored paradigm for the future of work with AI~\citep{acemoglu2026building}. While recent benchmarks have focused on evaluating fully autonomous agents on occupational tasks~\citep{patwardhan2025gdpval,mazeika2025remote,vidgen2025ai,vidgen2026apex}, effective human augmentation cannot be assumed to follow directly from strong autonomous performance as collaborative outcomes are shaped by many factors, including human AI literacy, the collaborative behaviors of models, and interaction design~\citep{shao2024collaborative,zou2025llm,mozannar2025magentic,haupt2025position}.

Benchmarking agents in collaborative settings, however, is fundamentally more complex than evaluating autonomous agents. %
A dominant paradigm is to use LLMs as user simulators that generate user turns to drive the interaction~\citep{yao2024tau,zhou2024sotopia,vijayvargiya2026interactive,zhou2025haicosystem,qian2025userbench}. While simulated users provide a useful proxy for scalable testing, research has shown that LLM simulators tend to be excessively cooperative, stylistically uniform, and devoid of realistic frustration~\citep{zhou2026mind,mehri2026measuring,suh2026quantifying}. Moreover, simulated users offer limited insight into how real humans engage with AI agents in their actual work or how they perceive them~\citep{hu2025simbench}. For studies that do engage real humans, they reveal more nuanced interaction patterns, but rely on simple tasks or focus narrowly on web navigation, failing to capture how professionals work day to day~\citep{shao2024collaborative, shen2025completion, huq2025cowpilot,drouin2024workarena}. As AI agents are increasingly deployed in economically valuable settings, this gap becomes consequential~\citep{meimandi2025measurement}.  To this end, we introduce \textbf{\name, a framework for evaluating human-agent collaboration on realistic, economically valuable tasks with real human workers}. \name provides \textbf{infrastructure for collecting human-agent collaboration data at scale} (\refsec{sec:data_infra}): a common task schema that links prompts, reference files, and deliverables to O*NET occupational categories~\citep{onet_database} supports heterogeneous tasks across sectors and allows workers to be matched to tasks aligned with their occupational background to mitigate variability in task familiarity; and a reference-free automated grader comprising rubric generation and multi-agent scoring handles open-ended deliverables. From the study period, we collected 386 sessions comprising over 1,500 user prompts across 10 O*NET sectors using this infrastructure.

Beyond data collection, a second challenge is that raw collaboration outcomes conflate agent and human performance. Workers vary substantially in AI literacy and agent familiarity, thus treating human participants as a factor to average over produces unreliable agent rankings. Rather than averaging across sessions from different humans, \name employs \textbf{a Bayesian system for disentangling and quantifying skill contributions from both human and AI agents} (\refsec{sec:ranking}). Inspired by multiplayer game rating~\citep{Herbrich2007TrueSkill,minka2018trueskill} that jointly estimates human and agent collaboration skill from team outcomes, the Bayesian skill rating system explicitly models inter-human variability in AI literacy as a latent human component rather than a source of noise. With the collected data points, it updates the posterior for both AI agents and human workers. The estimated human CollabSkill values also help surface insights into AI literacy across occupational sectors, complemented by pre- and post-task surveys capturing how workers perceive and experience AI agents in their professional workflows (\refsec{sec:survey}).

Our analysis yields insights on two fronts. \textbf{On the agent side, \name produces rankings that diverge from fully autonomous evaluations.} Among terminal-based agents which share the ReAct-based agent scaffolding and differ primarily in their underlying model, Claude Code ranks first (CollabSkill=74.8), a rank reversal relative to autonomous evaluations, where Codex consistently leads Claude Code across our own evaluation in the solo agent setting and public leaderboards like SWE-bench Verified, GDPVal, and SciCode~\citep{patwardhan2025gdpval,jimenez2024swebench,tian2024scicode}. %
Moreover, Claude Cowork ranks above Claude Code despite sharing the same underlying LLM, isolating interface design as a desiderata of collaboration agents across the broader spectrum of work beyond software engineering. Together, these results show that autonomous evaluation alone is insufficient for capturing human-agent collaboration. \textbf{On the human side, \name reveals that practical experience drives human CollabSkill score, and that hands-on collaboration shifts workers' AI literacy.} Practical LLM experience correlates with collaboration skill (Spearman $\rho=0.297, p=0.041$), whereas self-reported attitudes largely do not. Hands-on collaboration also shifts workers' perceived AI capability toward greater autonomy ($\Delta=-0.50$, $p<0.001$), yet preferred autonomy for meaningful work remains unchanged, suggesting that experience updates beliefs about what agents \textit{can} do without changing what workers \textit{want} them to do. Together, these findings call for agent development to focus on the unique demands of human-agent collaboration and suggest \name can serve as a framework for investigating both AI agents' collaboration capability and workers' AI literacy.

To sum up, we make the following contributions:
\begin{itemize}[itemsep=0pt]
    \vspace{-0.5em}
    \item {We introduce \name, a framework for evaluating human-agent collaboration on real-world tasks. \name employs a Bayesian skill rating system that explicitly models human collaboration skill to account for inter-human variability, yielding more reliable agent rankings and deeper insights into human worker behavior.}
    \item {Across 386 sessions in 10 O*NET sectors, \name yields agent rankings that diverge from fully autonomous benchmarks and enables comparison across agents with different LLM backbone, agent harness, and interface designs.}
    \item {By involving real human workers, \name contributes insights into AI literacy, revealing how practical experience, attitudes, and collaboration skill interrelate.%
    }
\end{itemize}

\section{Related Work}
\textbf{Human-Agent Collaboration}\quad
Effective human-agent collaboration has been shown to improve reliability~\citep{dong2025correctness,mozannar2025magentic}, enable more complex task completion~\citep{huq2025cowpilot}, and enhance outcome quality~\citep{shao2024collaborative,shen2025completion,luo2025hai} in comparison to full automation. These benefits stem from two underlying assumptions: human agency that humans have a desire for control and implicit requirements over task outcomes~\citep{shao2025future}; and complementary synergy, where humans and agents each bring distinct strengths~\citep{wang2025ai}.%

Despite this potential, systematic evaluation of human-agent collaboration remains lacking. On one hand, the HCI community has produced a proliferation of collaboration systems~\citep{feng2024cocoa,pu2025assistance,xu2025duetui,yao2025through}, yet these focus on system design rather than rigorous benchmarking. On the other hand, the dominant evaluation paradigm in the AI community uses LLM-simulated users~\citep{yao2024tau,zhou2024sotopia,vijayvargiya2026interactive,zhou2025haicosystem,qian2025userbench}, which enable scalable testing but tend to be excessively cooperative and offer limited insight into how real humans engage with agents~\citep{zhou2026mind,mehri2026measuring,suh2026quantifying}. Moreover, these studies do not account for humans' latent skill when evaluating agents in the human-agent collaboration setting,  which confounds the results as prior work reveals that human reliance level and individual success score can significantly affect human-AI collaboration outcome~\citep{guo2024decision,arnaiz2025towards,davidson2025collaboration}.

\textbf{Agent Ranking}\quad
Leaderboards and rankings are a driving force for AI advancement, typically following one of two approaches: absolute score-based ranking and comparison-based ranking. Absolute score-based ranking is simple to aggregate and widely adopted in agent leaderboards such as SWE-Bench~\citep{jimenez2024swebench} and OSWorld~\citep{xie2024osworld}. Comparison-based ranking is preferred when absolute scores are hard to obtain or tasks involve subjective preference~\citep{ammar2011ranking}. For example, GDPVal ranks models by average win-rate against expert-produced outcomes~\citep{patwardhan2025gdpval}, and Elo-style systems are widely used in both chatbot~\citep{bai2022training,chiang2024chatbot} and agent leaderboards~\citep{mazeika2025remote} when no specific anchor exists.  \name differs from standard leaderboards in that observed outcomes reflect the joint contribution of the agent and the human rather than the agent performance alone. This introduces additional sources of variation including differences in human skill and non-stationarity as users learn or fatigue over time. We therefore adopt a Bayesian skill rating system inspired by multiplayer game platforms like XBox Live~\citep{graepel2007bayesian,minka2018trueskill} that jointly estimates the skill of both humans and agents from team outcomes.

\textbf{The Future of Work with AI Agents}\quad
A broad body of work in digital economics has examined the implications of the recent surge of LLMs and AI agents~\citep{demirci2025ai,eloundou2024gpts,handa2025economic,hoffmann2025generative}, finding significant productivity opportunities alongside substantial economic and societal risks, including job loss, inequality, and reduced human agency~\citep{hazra2025position,brynjolfsson2025canaries}. Notably, AI augmentation of human workers already accounts for the majority of real-world AI use. \citet{handa2025economic} find that augmentation comprises 57\% of interactions versus 43\% for full automation in Claude usage. Yet this paradigm remains underexploited relative to its transformative potential~\citep{acemoglu2026building}. Prior work has approached the augmentation question through analyzing real user data~\citep{handa2025economic} and through large-scale worker surveys~\citep{shao2025future}. \name complements these efforts with a performance-based perspective.%

\begin{figure*}[t]
    \centering
    \resizebox{\textwidth}{!}{%
    \includegraphics{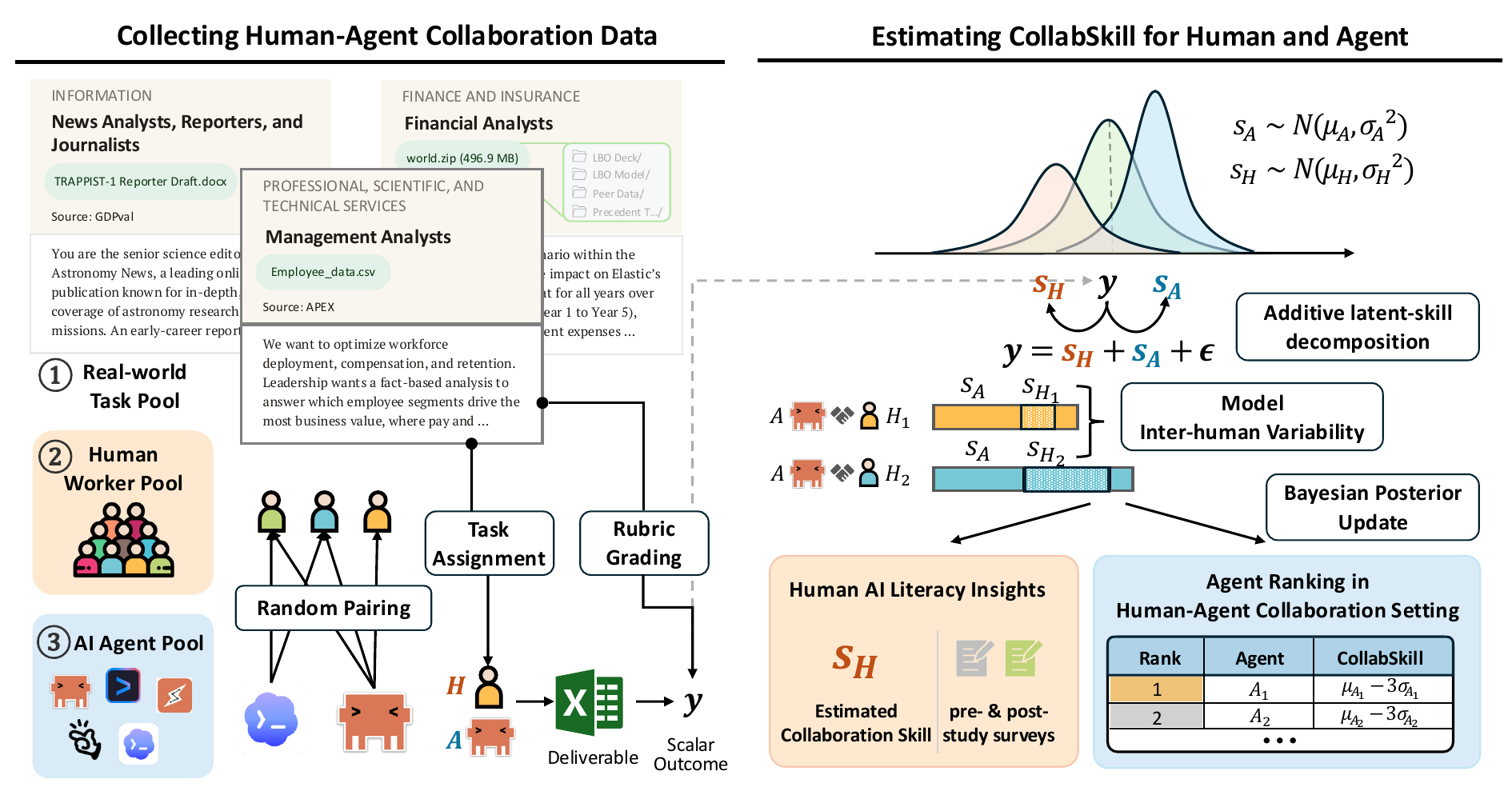}
    }
    \vspace{-1.5em}
    \caption{\textbf{Overview of \name.} \name provides infrastructure for collecting human-agent collaboration data at scale (Left) and a Bayesian skill rating framework to disentangle human and AI contributions (Right). Together, these yield agent rankings in human-agent collaboration setting and insights into human workers' AI literacy.%
    }
    \label{Fig:overview}
    \vspace{-1em}
\end{figure*}

\section{Human-Agent Collaboration Data Collection}
\label{sec:data_infra}
In this section, we discuss our infrastructure design for collecting human-agent collaboration data on occupational tasks at scale and present data collection statistics.

\subsection{Overview}
\label{sec:data_infra_overview}
A datapoint in \name is one completed evaluation episode in which a human works with an AI agent on a task instance under a fixed scoring function.%
\begin{itemize}[itemsep=0pt]
    \item {\textbf{Task ($t$):} \name targets realistic occupational tasks that professionals carry out in their jobs---assignments that require reference files, specialized software, and produce concrete deliverables (\reffig{Fig:overview} \Circled{1}). %
    We propose \texttt{TaskInstance}, a schema with a unique UUID comprising a natural-language prompt, hidden evaluator information (\eg, gold deliverables, domain knowledge), reference files, required software, expected deliverables. %
    Each \texttt{TaskInstance} is tagged with O*NET occupational metadata, %
    enabling matching users with tasks by their occupational background.
    }
    \item {\textbf{Human ($H$):} \name requires users to provide their occupational background, which is used to match them to appropriate tasks (\reffig{Fig:overview} \Circled{2}).}
    \item {\textbf{Agent ($A$):} \name supports any agent with an installation and usage guide. Our experiment includes %
    (\reffig{Fig:overview} \Circled{3}): (i) three terminal-based agents---Codex, Gemini CLI, and Claude Code---built on GPT, Gemini, and Claude backbones respectively; (ii) Claude Cowork, a desktop agent sharing the same implementation as Claude Code but with a graphical interface designed for non-developers; and (iii) Manus, a general-purpose browser-accessible agent.\footnote{The specific model versions used are \texttt{gpt-5.2-codex-medium}, \texttt{gemini-3-auto}, \texttt{claude-sonnet-4.6} (for both Claude Code and Claude Cowork), and \texttt{manus-1.6-lite}.} This selection enables controlled comparison across LLM backbones, agent harnesses, and interface designs.}
    \item {\textbf{Scalar Outcome ($y$):} Once assigned a task and paired with an agent, the user is instructed to set up the agent on their local machine, complete the task, and submit the delivered outcome on \name platform. \name then employs a reference-free automated grader (see \refsec{sec:autograder}) to assign a 0-100 scale score based on the task and submitted deliverables.}
\end{itemize}

\subsection{Grading Task Outcome}
\label{sec:autograder}
Occupational task deliverables are open-ended by nature, such as Excel workbooks, PDFs, slide decks,
ZIP archives, and audio files. %
To obtain a scalar outcome $y$ for each episode, \name employs a two-stage reference-free automated grading pipeline comprising
rubric generation followed by multi-agent scoring.

\textbf{Rubric Generation}
Given a task $t$ (comprising the task prompt, reference files, and other available information), we employ an agent\footnote{We use Gemini CLI in headless mode for rubric generation.} to produce a JSON rubric $\mathcal{R}(t)$ that partitions evaluation criteria into weighted top-level categories including
\textit{Correctness \& Accuracy}, \textit{Deliverable Completeness},
\textit{Task Requirements}, and \textit{Technical Quality} (see~\reffig{prompt:rubric_generation}). Each rubric item
specifies not only the criterion and its point value but also
the verification procedure (\eg, ``Test all links for accessibility
and verify they match described services'').

\textbf{Multi-agent Scoring}\quad
Given rubric $\mathcal{R}(t)$ and the submitted deliverables $d$, the grader invokes $N$ independent
agents $\{f_i\}_{i=1}^N$, each receiving $t, \mathcal{R}(t)$, and $d$, and returning a per-criterion score breakdown summing to a total $\hat{y_i}\in[0,100]$.\footnote{In our implementation, $N=2$ and we employ Gemini CLI and Claude Code as the judges.} The final score $y$ is the average of $\{\hat{y_i}\}_{i=1}^N$. This multi-agent design mitigates self-preference bias in LLM-as-a-judge evaluation, whereby a model acting as both
generator and evaluator systematically inflates scores for its own outputs~\citep{wataoka2024self}.

We validate the automated grading pipeline against official hand-authored GDPVal rubrics which were released after the original GDPval paper~\citep{patwardhan2025gdpval}. Manual rubric audit across 27 tasks across all 9 sectors from GDPval shows that our auto-generated rubrics achieve 82.1\% recall and 92.2\% precision against official criteria, with all generated items judged as important (see Appendix~\ref{appendix:autograder_validation} for details).%

\subsection{Data Collection Results}
\label{sec:data_collection_results}
\textbf{Task Sourcing}\quad
We unify three public datasets: (i) \textbf{GDPVal}~\citep{patwardhan2025gdpval}, 220 economically valuable tasks across 44 occupations; (ii) \textbf{APEX}~\citep{vidgen2025ai}, professional tasks across investment banking, management consulting, law, and primary care; and (iii) \textbf{APEX-Agents}~\citep{vidgen2026apex}, tasks across investment banking, law, and management consulting requiring cross-application execution with extensive reference files. Together, the sourced tasks cover 10 out of 20 O*NET work sectors (see Appendix~\ref{appendix:task} for task distribution details).

\begin{wraptable}{r}{0.3\textwidth}
\vspace{-1em}
\begin{minipage}{0.3\textwidth}
\caption{\textbf{Data Statistics.}%
}
\resizebox{\textwidth}{!}{%
\begin{tabular}{rl} 
\toprule
\multicolumn{1}{l}{\# Participants}            & 93   \\
\multicolumn{1}{l}{\# Data points}             & 386  \\
\multicolumn{1}{l}{\# Distinct Tasks Executed} & 165  \\
\midrule
\multicolumn{1}{l}{Avg. Years of Experience}   & 9.6  \\
0-2 years                                      & 15   \\
3-5 years                                      & 32   \\
6-10 years                                     & 17   \\
11-20 years                                    & 17   \\
20+ years                                      & 10   \\
Not Reported                                   & 2    \\
\bottomrule
\end{tabular}

}
\label{table:data_statistics}
\end{minipage}
\vspace{-1em}
\end{wraptable}

\textbf{Human Worker Recruitment}\quad
We recruit U.S.-based participants via Upwork, compensating \$20 per completed task with a suggested five-task target though they may withdraw early at their discretion. Tasks are randomly matched to participants based on their occupational background.  To incentivize participants to complete these tasks carefully, those ranked 1st, 2nd, and 3rd by CollabSkill score will be awarded bonuses of \$100, \$50, and \$50, respectively. Before starting each session, participants are randomly paired with an AI agent, presented with the installation guide for their assigned agent, and instructed to watch a tutorial video if they have not used it before. They then proceed to the task on the platform, where the task prompt and any reference files are provided. Upon completion, participants submit their final deliverables along with an interaction log (screenshots or terminal content or shareable links for Manus). 
Participants completing all five tasks are invited to fill out a post-study survey. The study protocol is IRB-approved. Interface details are in Appendix~\ref{appendix:interface}.

\textbf{Data Statistics}\quad
During the study period, 93 human workers participate in \name, contributing 386 sessions across 10 O*NET sectors. 76 participants complete both pre/post- study surveys. Participants have an average of 9.6 years of professional experience in their selected occupation; detailed statistics are provided in \reftab{table:data_statistics}.

\section{From Teamwork Outcomes to CollabSkill Score}
\label{sec:ranking}

The scalar score in each collected data point reflects the joint contribution of both the human and the agent. Simply averaging scores per agent cannot produce a trustworthy ranking due to inter-human variability as people could differ substantially in AI literacy.  %
In this section, we introduce the Bayesian skill rating system employed by \name to disentangle and estimate the collaboration skill from both sides, which draws inspiration from ranking methods in multiplayer video games~\citep{Herbrich2007TrueSkill,minka2018trueskill}.

Specifically, for each teamwork outcome observation $(A, H, y)$, prior work typically estimates agent quality as the sample mean across sessions, $\hat{q}_A = \frac{1}{|\mathcal{D}_A|}\sum_{(A,H,y)\in \mathcal{D}_A} y$, where $\mathcal{D}_A$ denotes the set of sessions involving agent $A$, implicitly assuming that human variability averages out in expectation. Rather than making this assumption, we propose to design latent skill values $s_A$ and $s_H$ to the agent and human respectively, and model the scalar score $y$ as an additive decomposition of agent skill, human skill, and observation noise. 
\begin{equation}
    y = s_A + s_H + \epsilon, \qquad \epsilon \sim \mathcal{N}(0,\beta^2).
    \label{eq:model}
\end{equation}

This formulation is the key modeling choice that separates \name from prior metrics such as task completion rate, average session score, and win rate: rather than treating human variability as noise, Equation (\ref{eq:model}) explicitly allocates a latent skill component to each human participant. For example, when agent $A$ collaborates with both $H_1$ and $H_2$, the model estimates $s_{H_1}$ and $s_{H_2}$ separately from their respective sessions; the agent's skill $s_A$ is then inferred by explaining away the portion of each outcome attributable to the human. Intuitively, if $H_1$ consistently scores well across multiple agents while $H_2$
does not, the model attributes that difference to human skill in collaborating with AI agents rather than to whichever agent $H_1$ happened to work with.

Each entity is initialized with a Gaussian prior. For an agent $A$ and a human user $H$,
\begin{equation}
    s_A \sim \mathcal{N}(\mu_{A,0}, \sigma^2_{A,0}), 
\qquad
s_H \sim \mathcal{N}(\mu_{H,0}, \sigma^2_{H,0}).
\label{eq:prior}
\end{equation}
When a new observation $(A, H, y)$ arrives, consider the two-dimensional latent state $\theta = [s_A, s_H]^\top$ with prior $\theta \sim \mathcal{N}(m,\Sigma)$. Writing the design vector as $x=[1,1]^\top$, the observation takes the form
$ y = x^\top \theta + \epsilon $.
Since this is a linear-Gaussian model, the posterior $N(m', \Sigma')$ is available in closed form via the Kalman filter measurement update~\citep{Kalman1960} (see Appendix~\ref{appendix:deriving_update} for derivation):
\begin{align}
    r = y - x^\top m,\qquad
K = \frac{\Sigma x}{x^\top \Sigma x + \beta^2},\\
m' = m + Kr,\qquad
\Sigma' = \Sigma - Kx^\top \Sigma,
\label{eq:posteria_update}
\end{align}
where $r$ is the prediction error (how much the outcome surprised the model) and $K$ is the Kalman gain (how much weight to place on the new observation relative to the prior). When an outcome exceeds expectations, both $\mu_A$ and $\mu_H$ increase, with larger adjustments assigned to the entity with greater uncertainty $\sigma^2$. As observations accumulate, $\sigma^2$ shrinks and estimates stabilize. 

Crucially, the model does not require a large number of sessions per agent to produce reliable rankings. What matters is the connectivity of the human-agent interaction graph, \ie, every agent is linked to every other agent through at least one shared human participant, and vice versa. When the same human $H$ appears in sessions with both agent $A_1$ and agent $A_2$, their shared outcomes allow the model to attribute score differences to the agents rather than to $H$ alone, propagating information across the graph.  In \name, where a small fixed set of agents is evaluated and humans are randomly matched to available agents, such overlap arises naturally and guarantees sufficient connectivity for reliable estimation.

In practice, we initialize all entities with $\mu_0=0$ and $\sigma_0=1$, and estimate the posterior efficiently using the information form of the Gaussian (see Appendix~\ref{appendix:estimating_var}). \name defines \textbf{CollabSkill score} $\triangleq \mu_i - 3\sigma_i$ as a conservative measure of collaboration capability (approximately the 1\% lower quantile under a Gaussian) that penalizes high-uncertainty entries and prevents over-ranking agents with few observations~\citep{Herbrich2007TrueSkill}. CollabSkill scores are meaningful when compared within the same group---a higher score indicates that the entity is likely to contribute more to human-agent collaboration outcomes. In Appendix~\ref{appendix:rating_method_comparison}, we further show empirically that the CollabSkill rating system yields lower variance when comparing agents than naively averaging performance.

\section{Human Factors: AI Literacy and Attitudes}
\label{sec:survey}
Involving real humans in \name lets us examine how AI literacy shapes human-agent collaboration and how it evolves through the experience. To complement CollabSkill of the user, we further collect attitudinal readiness, measured through paired pre- and post-task surveys designed around the Technology Acceptance Model ~\citep{davis1989perceived}. This design lets us ask questions with direct practical stakes: (i) Which prior experience and attitudinal factors best predict collaboration skill? (ii) Does hands-on collaboration recalibrate workers' beliefs about agent capability or shift their preferred autonomy levels?

\textbf{External Variables: Prior Exposure and Agent Familiarity}\quad
In the demographics survey administered at the start of the study, we capture each participant's baseline familiarity with LLMs (\ref{item:D1}) and their frequency of LLM use in professional contexts (\ref{item:D2}). We further ask participants to rate their familiarity with AI agents defined as ``systems that autonomously complete tasks on your behalf'' (\ref{item:A1}).

\textbf{Perceived Usefulness: Trust and Delegation Comfort}\quad
The Technology Acceptance Model posits that readiness to adopt a technology is strongly predicted by perceived usefulness. We measure this construct through two complementary instruments: (i) \textit{trust in agent autonomy}, participants rate how much they trust AI agents to complete tasks accurately without human oversight, capturing perceived reliability (\ref{item:A2}); and (ii) \textit{comfort with delegation}, participants rate how comfortable they are delegating work tasks to an AI agent (\ref{item:A3}).

\textbf{Perceived Capability: Autonomy Calibration}\quad
We assess how participants perceive human-agent autonomy sharing. As AI agent use is not a binary decision but a spectrum, we adopt the Human Agency Scale (HAS)~\citep{shao2025future}, a shared framework for quantifying automation versus augmentation, which defines five levels from full agent autonomy (\textit{H1}), minimal human input (\textit{H2}), equal partnership (\textit{H3}) to human lead (\textit{H4}) and full human involvement (\textit{H5}) (see \reftab{table:has}). Participants identify the level at which they \textit{believe} AI agents can currently operate for their typical work tasks (\ref{item:A4}).

\textbf{Attitude: Desired Autonomy and Affective Orientation}\quad
Using the same HAS, participants indicate their \textit{preferred} autonomy level (\ref{item:A5}), their concern that AI will reduce their role below what they consider meaningful (\ref{item:A6}), and their overall affective orientation toward AI agents, chosen from \textit{Excited}, \textit{Curious}, \textit{Neutral}, \textit{Skeptical}, or \textit{Anxious} (\ref{item:A7}).

\textbf{Pre/Post Paired Design}\quad
All items (except demographics and LLM familiarity) are administered identically before and after participants complete a series of collaborative tasks with AI agents. %
This paired design allows us to examine how hands-on experience with AI agents in occupational tasks shapes participants' AI literacy and attitudes.

\section{Analysis}
\begin{table*}[t]
\centering
\begin{minipage}[t]{0.6\textwidth}
    \centering
    \captionof{table}{\textbf{Agent rankings on \name.} \name estimates each agent's latent skill as a Gaussian and reports its posterior mean $\mu$ and standard deviation $\sigma$. Agents are ranked by the conservative CollabSkill $\mu - 3\sigma$, following ~\citet{Herbrich2007TrueSkill}, to penalize highly uncertain estimates.}
    \resizebox{\textwidth}{!}{%
        \begin{tabular}{lccc} 
\toprule
\multicolumn{1}{c}{\textbf{Agent}} & \multicolumn{1}{c}{\textbf{CollabSkill}} & \multicolumn{1}{c}{$\mu$} & \multicolumn{1}{c}{$\sigma$}   \\ 
\midrule
Claude Cowork \footnotesize{\texttt{claude-sonnet-4.6}}  & 76.738                                 & 77.206                    & 0.156                                                  \\
Claude Code \footnotesize{\texttt{claude-sonnet-4.6}}    & 74.778                                 & 75.248                    & 0.157                                           \\
Codex \footnotesize{\texttt{gpt-5.2-codex-medium}}       & 71.267                                 & 71.733                    & 0.155                                          \\
Manus \footnotesize{\texttt{manus-1.6-lite}}             & 71.204                                 & 71.646                    & 0.147                                        \\
Gemini CLI \footnotesize{\texttt{gemini-3-auto}}         & 69.555                                 & 70.009                    & 0.151                                          \\
\bottomrule
\end{tabular}

    }
    \label{table:ranking}
\end{minipage}
\hfill
\begin{minipage}[t]{0.37\textwidth}
    \centering
    \captionof{table}{\textbf{CollabSkill ranking stability under bootstrap resampling ($N=10,000$).} We report each agent's mean rank and its probability of ranking first across $N$ resampling rounds.}
    \resizebox{\textwidth}{!}{%
        \begin{tabular}{lcc} 
\toprule
\multicolumn{1}{c}{\textbf{Agent}} & \multicolumn{1}{c}{\textbf{Mean}} & \multicolumn{1}{c}{\textbf{P(rank=1)}}  \\ 
\midrule
Claude Cowork                      & 1.36                              & 0.713                                   \\
Claude Code                        & 2.08                              & 0.233                                   \\
Codex                              & 3.58                              & 0.036                                   \\
Manus                              & 3.64                              & 0.014                                   \\
Gemini CLI                         & 4.34                              & 0.004                                   \\
\bottomrule
\end{tabular}

    }
    \label{table:ranking_stability}
\end{minipage}
\end{table*}

\begin{table}[t]
\centering
\caption{\textbf{\name ranking stability under leave-one-out tests.} We report mean agent rankings under two leave-one-out conditions: dropping all sessions from one human at a time, and dropping all sessions from one \texttt{TaskInstance} at a time.}
\resizebox{0.85\textwidth}{!}{%
    \begin{tabular}{lccc} 
\toprule
\multicolumn{1}{c}{\textbf{Agent}} & \textbf{Ranking} & \textbf{Leave-One-Human-Out Mean} & \textbf{Leave-One-Task-Out Mean}  \\ 
\midrule
Claude Cowork                      & \#1              & 1.00                              & 1.00                              \\
Claude Code                        & \#2              & 2.00                              & 2.00                              \\
Codex                              & \#3              & 3.41                              & 3.33                              \\
Manus                              & \#4              & 3.59                              & 3.67                              \\
Gemini CLI                         & \#5              & 5.00                              & 5.00                              \\
\bottomrule
\end{tabular}

}
\vspace{-1em}
\label{table:loho_loto}
\end{table}

\begin{figure}[t]
    \centering
    \begin{minipage}[c]{0.55\textwidth}
        \centering
        \captionof{table}{\textbf{Comparison of CollabSkill and solo agent performance.} %
        We run terminal-based agents in headless mode with an AI researcher-engineered system prompt on 165 collected tasks and report the mean score under Solo Score. SWE-bench Verified, GDPval, SciCode results are sourced from their official leaderboards. Model versions match those used in \refsec{sec:data_infra_overview}.}
        \resizebox{\textwidth}{!}{%
        \begin{tabular}{lc!{\vrule width \lightrulewidth}cccc} 
\toprule
\multicolumn{1}{c}{\textbf{Agent}} & \textbf{CollabSkill} & \begin{tabular}[c]{@{}c@{}}\textbf{Solo}\\\textbf{Score}\end{tabular} & \begin{tabular}[c]{@{}c@{}}\textbf{SWE-bench}\\\textbf{Verified}\end{tabular} & \textbf{GDPval} & \textbf{SciCode} \\ 
\midrule
Claude Code                        & 74.778               & 77.17                                                                            & 71.40                                                                         & 42.5\%    & 46.8       \\
Codex                              & 71.267               & 78.02                                                                            & 72.80                                                                         & 49.7\%     & 54.6      \\
Gemini CLI                         & 69.555               & 79.44                                                                            & 69.60                                                                         & 40.3\%   & 56.1         \\
\bottomrule
\end{tabular}

        }
    
        \label{table:autonomous_results}
    \end{minipage}
    \hfill
    \begin{minipage}[c]{0.42\textwidth}
        \centering
        \includegraphics[width=\textwidth]{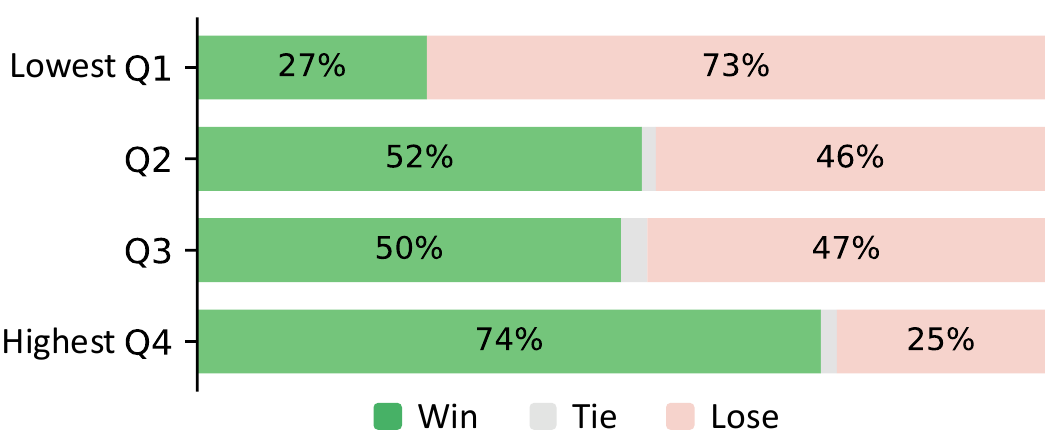}
        \vspace{-2em}
        \captionof{figure}{\textbf{Human-agent collaboration vs.\ solo agent win rates.} Pairwise comparison between human-agent collaboration sessions and solo agent counterparts with an AI researcher-engineered system prompt grouped by estimated human CollabSkill score.}
        \label{Fig:hac_autonomous_comparison}
    \end{minipage}
    \vspace{-1em}
\end{figure}

\subsection{The Agent Side in Human-Agent Collaboration}

\reftab{table:ranking} reports agent rankings on \name. Among the five agents evaluated, Claude Cowork achieves the highest estimated CollabSkill ($\mu-3\sigma$) as 76.738. Since our experiment includes agents with different LLM backbones, agent harness, and interface designs, we structure the comparison along two axes. 

\textbf{Axis 1: Model Collaboration Strength Among Terminal-based Agents}\quad
Codex, Claude Code, and Gemini CLI employ a similar terminal-based interface, and are all ReAct-based agents with bash tools, so their relative ranking primarily reflects differences in the underlying models. Among them, Claude Code performs best in the human-agent collaboration setting ($p(s_{\text{Claude Code}} > s_{\text{Codex}}) > 0.999$)\footnote{Since the Bayesian skill rating system models the each agent's skill as a Gaussian, the difference is also Gaussian. We compute this probability $p$ using the CDF.} ---a ranking that diverges from both our autonomous evaluation and the rankings reported by SWE-bench Verified, GDPVal, and SciCode (\reftab{table:autonomous_results}), suggesting that model strength in autonomous settings does not straightforwardly transfer to collaboration.

\textbf{Axis 2: The Effect of Human-Agent Interface Design}\quad
Second, Claude Cowork and Claude Code share the same underlying LLM, so their comparison isolates the effect of interface design. Claude Cowork ranks strictly above Claude Code ($p(s_{\text{Claude Cowork}} > s_{\text{Claude Code}}) > 0.999$), demonstrating that interface-based agents are more favorable than terminal-based ones across the broader spectrum of human work beyond software engineering. Note that this comparison controls for installation friction, as we provided detailed installation guides and assisted participants who got stuck at the installation stage (see \reffig{fig:interface-setup}).  To further control for prior agent familiarity since interface-based agent may be easier to get started with, we recompute CollabSkill ratings restricting to participants with self-reported agent familiarity $\geq4$ on a 7-point Likert scale ($N=81$); the comparison between Claude Cowork and Claude Code remains unchanged (Claude Cowork: $\text{CollabSkill}=76.403, \sigma=0.173$; Claude Code: $\text{CollabSkill}=74.349, \sigma=0.175$).

\textbf{Validating \name Ranking}\quad
To assess the stability of our Bayesian skill rating system, we perform bootstrap resampling by drawing observations with replacement and recomputing the full posterior skill estimates across 10,000 rounds. As shown in \reftab{table:ranking_stability}, the top-ranked agent (Claude Cowork) retained rank 1 in 71.3\% of resamples (mean rank 1.36). The overall ranking order remained moderately stable (mean Kendall's $\tau= 0.64$), with most variation arising from adjacent-rank swaps between Codex and Manus.

We additionally conduct leave-one-human-out and leave-one-task-out analyses by recomputing rankings after removing each human annotator or each task in turn to check if any single human annotator or task disproportionately influences the overall ranking. Table~\ref{table:loho_loto} reports the mean ranking across all such recomputations. The results confirm that top two agents (Claude Cowork and Claude Code) and the bottom agent (Gemini CLI) receive perfectly stable mean rankings of 1.00, 2.00, and 5.00, respectively, under both analyses. The middle positions show minor variance: Codex and Manus swap slightly in mean rank between the two analyses (3.41 vs.\ 3.59 and 3.33 vs.\ 3.67, respectively), but their relative ordering remains consistent. Overall, these analyses indicate that the CollabSkill ranking is stable and not an artifact of any particular annotator or task.

\subsection{The Human Side in Human-Agent Collaboration}
We conduct head-to-head comparisons between human-agent teams and the same agent running autonomously under an AI-researcher-engineered system prompt (\reffig{prompt:autonomous_agent}). As shown in \reffig{Fig:hac_autonomous_comparison}, outcomes depend strongly on human worker CollabSkill score: top-quartile workers (Q4) achieve a 74\% win rate against the autonomous baseline, while bottom-quartile workers (Q1) win only 27\% of the time.%

\begin{wrapfigure}{r}{0.58\textwidth}
    \centering
    \includegraphics[width=\linewidth]{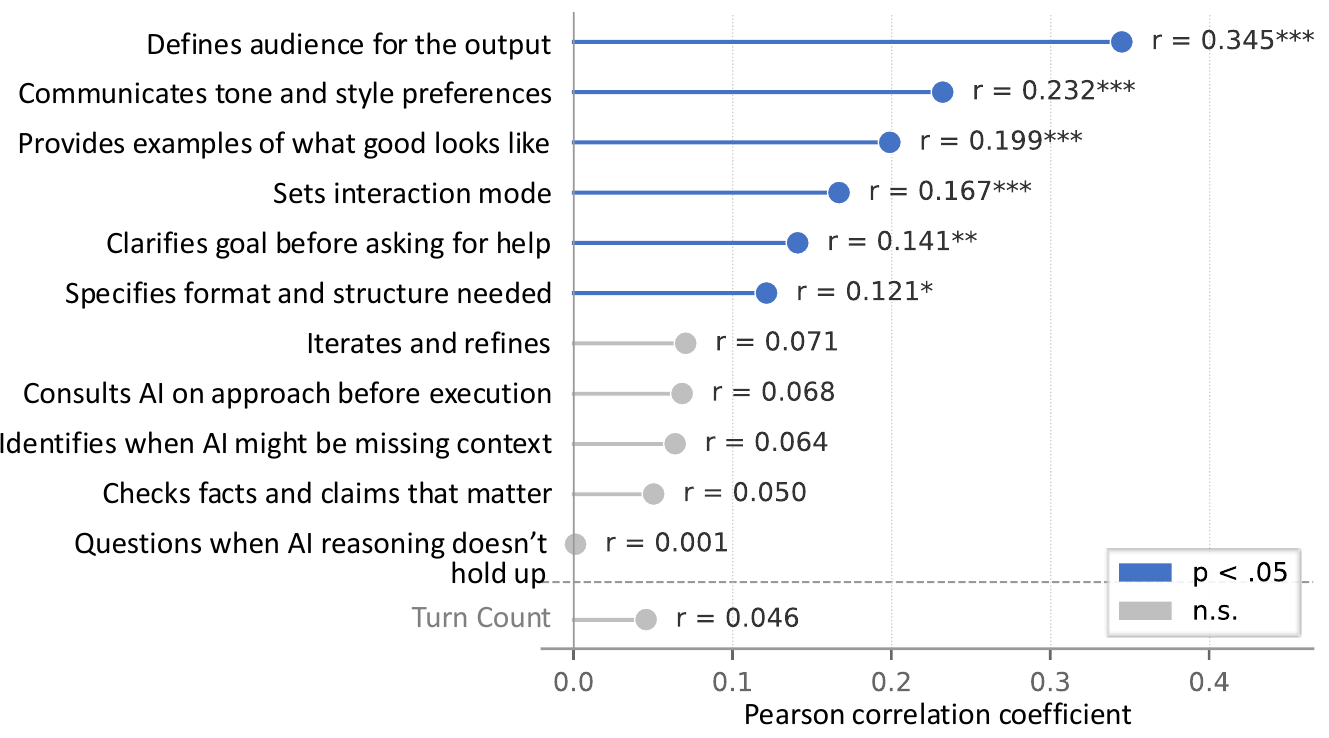}
    \caption{\textbf{Correlation between AI fluency behavior indicator with session score.}}
    \label{Fig:ai_fluency_correlation}
\end{wrapfigure}

\textbf{AI fluency behavior indicator predicts human-agent collaboration outcome}\quad
We analyze interaction logs by coding behaviors from the Anthropic AI Fluency Index which identifies 11 directly observable indicators of human skill in using AI~\citep{anthropic2026aifluency} (Appendix~\ref{sec:trajectory_analysis}). 
The total count of observed behaviors correlates strongly with session score (Pearson $r=0.262$, $p<1e-4$), with breakdowns shown in \reffig{Fig:ai_fluency_correlation}; by contrast, user turn count shows no significant correlation ($r=0.046$, $p=0.371$), suggesting that \textit{how} workers interact with agents matters more than \textit{how much}.

\begin{figure*}[t]
    \centering
    \resizebox{\textwidth}{!}{%
    \includegraphics{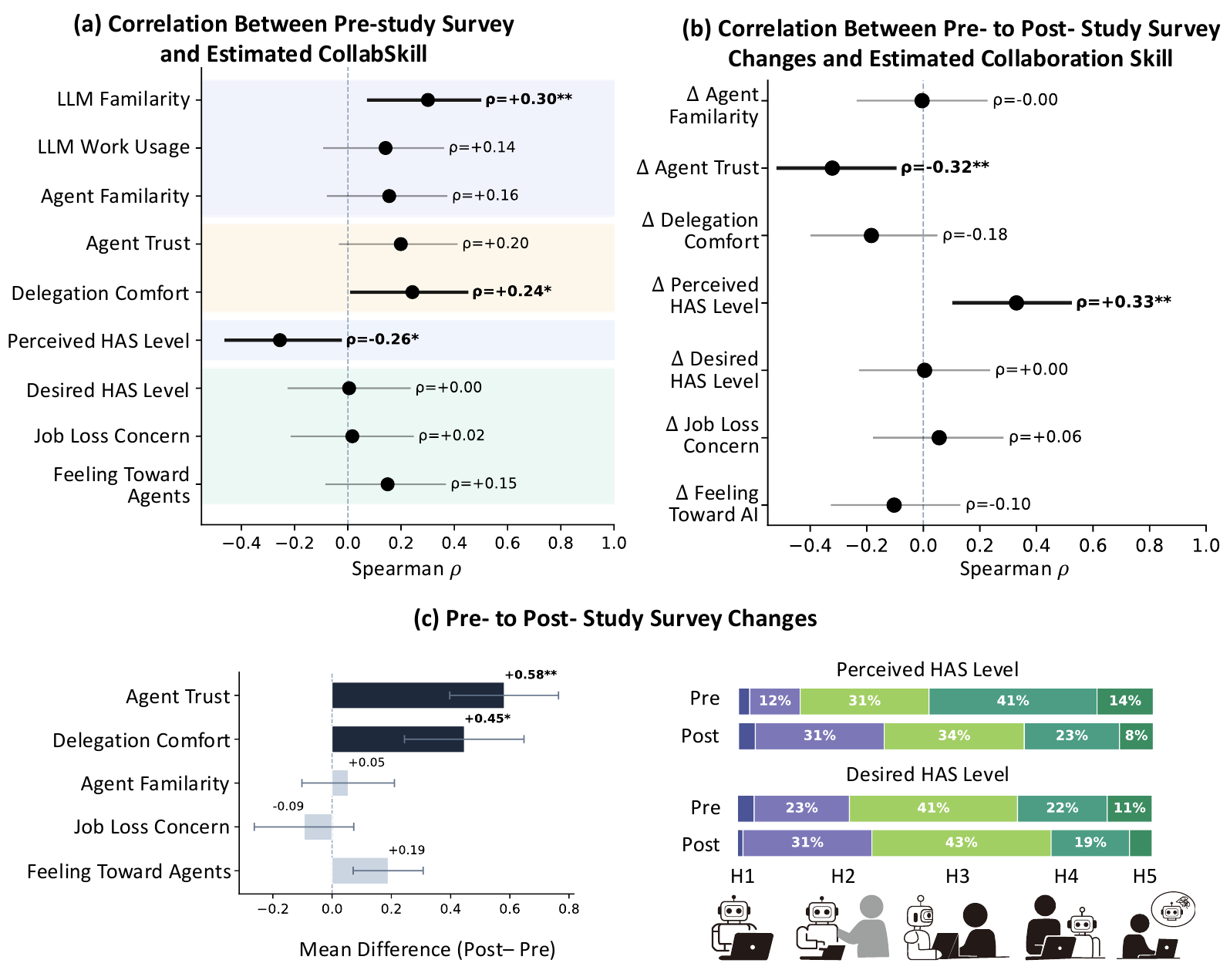}
    }
    \vspace{-1.5em}
    \caption{\textbf{Human CollabSkill and survey responses ($N=76$).} (a) Spearman rank correlations between pre-study survey measures and estimated CollabSkill; filled circles denote point estimates and horizontal bars show 95\% confidence intervals. (b) Spearman rank correlations between pre-to-post perception change and estimated CollabSkill. (c) Changes in perception after collaborating with AI agents on realistic occupational tasks; error bars denote standard errors computed via the Wilcoxon signed-rank test. In all three figures, * denotes $p<0.05$ and ** denotes $p<0.01$.
    }
    \vspace{-1em}
    \label{Fig:human_insights}
\end{figure*}

\textbf{Practical LLM experience predicts human's skill in collaborating with AI agents, but attitudes largely do not.}\quad
\name yields both self-reported attitudinal data and exhibited CollabSkill estimated by the Bayesian model. %
As shown in \reffig{Fig:human_insights} (a), among AI literacy measure collected in the pre-study survey and the participant's CollabSkill score estimate, self-reported LLM familiarity showed a significant positive correlation with collaboration skill (Spearman $\rho=0.297, p=0.010$).
Comfort with delegating work to an AI agent also correlated positively with skill ($\rho=0.238, p=0.041$). In contrast, none of the remaining attitudinal measures reached significance.

\textbf{Collaboration shifts perceived AI capability toward greater autonomy, but does not change the desired HAS level.}\quad
\reffig{Fig:human_insights} (c) reports paired pre- and post-study attitude changes with Wilcoxon signed-rank test. Participants reported increased trust ($\Delta=0.58, p=0.004$) and delegation comfort ($\Delta=0.45, p=0.039$). Notably, while participants revised their beliefs about AI capability toward greater autonomy (HAS belief $\Delta=-0.50, p<0.001$), the preferred autonomy level did not change ($\Delta=-0.16, p=0.105$), suggesting preferences about human-AI work allocation are governed by factors beyond perceived capability. When splitting participants at the median CollabSkill, participants in the lower half increased their agent trust by +1.03 points on average, compared to +0.16 for higher-skilled participants ($p=0.028$ in the Mann-Whitney U test), and shifted their perceived AI autonomy level by  -0.89 toward greater autonomy, versus -0.13 for higher-skilled participants ($p=0.001$). Spearman correlations corroborate this pattern as shown in \reffig{Fig:human_insights} (b).

\section{Conclusion}
This work presents \name, a framework for systematically evaluating human-agent collaboration on real-world occupational tasks. \name provides infrastructure for collecting collaboration data at scale and a Bayesian skill rating framework to estimate the collaboration skill of both human workers and AI agents. Our analysis yields practical findings for multiple stakeholders: for \textit{agent developers}, human-agent collaboration requires targeted evaluation and optimization, as autonomous capability alone does not determine collaboration quality; for \textit{employers}, workers' familiarity with AI agents significantly predicts their collaboration skill, and effective human-agent collaboration can surpass AI-only performance; for \textit{human workers}, hands-on experience collaborating with AI agents on occupational tasks is a meaningful pathway to improving AI literacy. We hope \name spurs efforts to build AI agents that augment human workers and to support workers who are genuinely ready to work with them.

\section*{Ethics Statement}

This study was approved by our Institutional Review Board. All 93 participants were recruited through Upwork on a voluntary basis and could withdraw at any point without penalty. All collected data are reported in anonymized and aggregate form.

Our Bayesian skill rating system estimates a latent collaboration skill $s_H$ for each participant strictly as a proxy for AI literacy within the human-agent collaboration setting we study. Individual estimates are used solely for the analyses presented in this paper. We caution against using estimated human skill scores to screen or evaluate individual workers for employment or other consequential decisions.

By focusing on human-agent collaboration, the work helps identify the strengths and weaknesses of AI agents in augmenting humans on occupational tasks, and supports the design of more effective collaborative agents. More broadly, it aligns with the goal of developing AI systems that preserve human agency and meaningful human control.

\section*{Acknowledgments}
We thank Caleb Ziems, Yanzhe Zhang, Chenglei Si, and Abe Hou for their valuable feedback on the manuscript, Haowen Wang and Boom Iamphongsai for testing the CollabSkill data collection interface, and all members of the Stanford SALT lab for their suggestions throughout this project. We are also grateful to faculty administrator Maria David for her dedicated assistance that makes this large-scale human evaluation possible. We extend special thanks to the human workers who participated in our study. Many of them generously offered voluntary comments on how they think about the future of work with AI agents and their experience with using AI agents in our study, which inspired us to dedicate a separate subsection to this topic. This research is supported in part by grants from Open Philanthropy, ONR N000142412532, NSF IIS 2247357, Schmidt Sciences, and a multi-company research collaboration via Stanford HAI with SCBx, Itau, Wells Fargo, and American Express.

\bibliography{colm2026_conference}
\bibliographystyle{colm2026_conference}

\clearpage
\appendix
\section{Disclosure of LLM use in Research}    
  We used LLMs in three parts of this work. First, during development of the \name frameworks and analysis   
  code, we used LLM-based coding agents and tab completion features to help write and debug code. Second, in preparing this 
  manuscript, we used LLMs to polish paragraphs originally drafted by the authors and to convert handwritten mathematical     
  derivations into LaTeX format. In both cases the authors wrote the initial content and verified the final output. Third,
  LLMs serve as components of the \name system itself: a coding agent generates evaluation rubrics from task specifications,
   and two LLM judges score submitted deliverables against those rubrics, as described in \refsec{sec:autograder}. %

\section{Limitations and Future Work}
While \name offers the first systematic framework for evaluating human-agent collaboration on realistic occupational tasks, several limitations should be considered.

\textbf{Evaluation Design.}\quad
Involving real human workers is central to \name's design: unlike agent-only benchmarks, \name captures authentic human–agent collaboration dynamics, and cross-analyzing estimated CollabSkill with self-reported survey results yields meaningful insights into workers' AI literacy. That said, our participant pool was recruited through Upwork and targets US-based workers only, so findings on human workers' CollabSkill scores and AI literacy should be interpreted with this selection bias in mind. With 386 sessions from 93 workers spanning 10 of the 20 O*NET sectors, \name establishes rankings and surfaces broader trends than most prior work focused exclusively on software engineering; however, we cannot conduct fine-grained subgroup analyses due to limited statistical power, and our sample is not yet representative of the full occupational landscape. Owing to the scalable \texttt{TaskInstance} schema, we are expanding \name to additional sectors and have released tooling on our open-source platform to support community contributions.

Simulation-based evaluation is another appealing direction, but we find current user simulators cannot be applied in our setting. Unlike typical chatbot scenarios where user simulators model turn-based dialogue, our setting involves professional workers completing complex, real-world occupational tasks through rich, multimodal interactions spanning agent interfaces (terminal, desktop, web app) and direct task-completion actions. The scale and naturalistic complexity of our human data reflects this: 386 sessions with a median duration of 76.7 minutes, yielding over 1,500 prompts and various editing actions in the task environment. We see building better user simulators for the human-agent co-working setting as a high-value research direction.

For task scoring, we use LLM judges rather than human graders. While human evaluation is the gold standard, it is prohibitively expensive at our scale. For example, GDPval~\citep{patwardhan2025gdpval} required substantial human grading costs and later introduced LLM-judge rubrics. A core design goal of \name is to remain accessible and reproducible for the broader research community, which necessitates automated scoring. While we've validated the automated grader (see Appendix~\ref{appendix:autograder_validation} for details), we acknowledge that improving automated graders is an important direction for future work.

\textbf{CollabSkill Formulation.}\quad
We model task outcomes as an additive decomposition of agent skill, human skill, and observation noise. Our primary objective is to disentangle human and agent skill and to provide robust quantitative estimates of each. A limitation of the current formulation is that the estimated CollabSkill score does not separately identify task skill and collaboration skill. We adopt this formulation because real-world tasks are substantially more complex than toy collaboration environments, making a clean decomposition challenging. Thus, we view the current formulation as a principled starting point; extending the CollabSkill ranking system to distinguish finer-grained factors is a promising direction for future work. Analogously, the original TrueSkill~\citep{Herbrich2007TrueSkill} ranks holistic team-play performance, and later extensions incorporate additional factors such as player experience and individual statistics to tailor rankings to specific games~\citep{minka2018trueskill}.

\section{\name Data Collection Details}

\subsection{Task Sourcing}
\label{appendix:task}
CollabSkill targets human-agent collaboration on real-world tasks. Our unified \texttt{TaskInstance} schema specifies both the task description (prompt) and execution environment (reference files, required software). For the experiments in this paper, tasks are sourced from three public datasets, GDPval~\citep{patwardhan2025gdpval}, APEX~\citep{vidgen2025ai}, and APEX-Agents~\citep{vidgen2026apex}, grounding task difficulty consistently across domains. We target a wide coverage of sectors and select 5 tasks per occupation from each source:
\begin{itemize}
    \item {\textbf{GDPval}: (1) Administrative Services Managers; (2) Buyers and Purchasing Agents; (3) Child, Family, and School Social Workers; (4) Compliance Officers; (5) Computer and Information Systems Managers; (6) Concierges; (7) Customer Service Representatives; (8) Editors; (9) Film and Video Editors; (10) Financial Managers; (11) Financial and Investment Analysts; (12) First-Line Supervisors of Office and Administrative Support Workers; (13) First-Line Supervisors of Police and Detectives; (14) General and Operations Managers; (15) Industrial Engineers; (16) Medical Secretaries and Administrative Assistants; (17) News Analysts, Reporters, and Journalists; (18) Nurse Practitioners; (19) Order Clerks; (20) Pharmacists; (21) Project Management Specialists; (22) Property, Real Estate, and Community Association Managers; (23) Real Estate Sales Agents; (24) Recreation Workers; (25) Sales Managers; (26) Shipping, Receiving, and Inventory Clerks; (27) Software Developers}
    \item {\textbf{APEX}: (1) Financial Analysts; (2) Healthcare Diagnosing or Treating Practitioners, All Other; (3) Management Analysts}
    \item {\textbf{APEX-Agents}: (1) Financial Analysts; (2) Lawyers; (3) Management Analysts}
\end{itemize}

Together, these tasks span 10 O*NET sectors, providing a diverse testbed for evaluating general-purpose agent capabilities in real-world human-agent collaboration settings. It is also worth noticing that the unified \texttt{TaskInstance} schema provides a standardized interface for incorporating new tasks. Since the initial version of this paper in March 2026, we have further integrated datasets such as JobBench\footnote{\url{https://huggingface.co/datasets/JobBench/job-bench}}, OccuBench\footnote{\url{https://huggingface.co/datasets/gregH/OccuBench}}, WorkspaceBench\footnote{\url{https://huggingface.co/datasets/Workspace-Bench/Workspace-Bench}}, ClawBench{\footnote{\url{https://huggingface.co/datasets/NAIL-Group/ClawBench}}} into our data pool.

 \subsection{\name Data Collection Interface}
  \label{appendix:interface}
\begin{center}
  \begin{figure}[h!]
      \centering
      \includegraphics[width=0.85\textwidth]{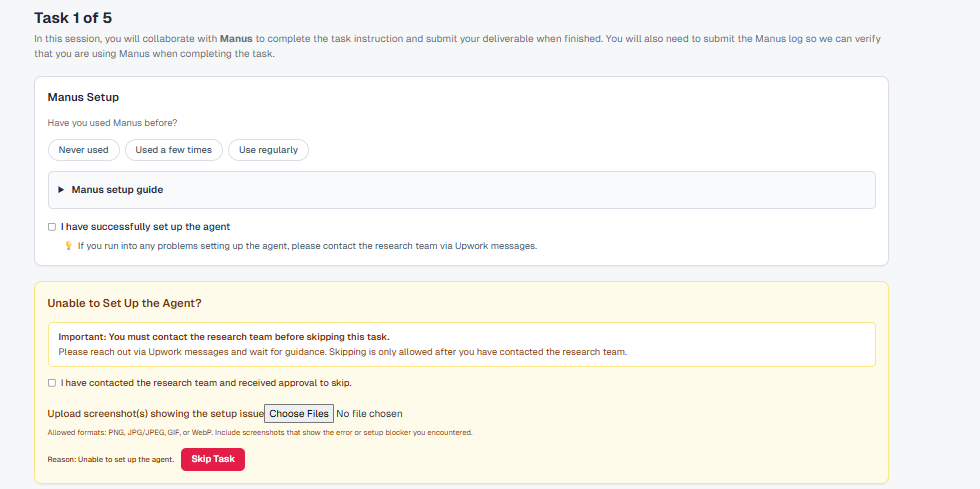}   
      \caption{\textbf{Agent setup panel on the task page.} The
  participant reports prior experience with the assigned agent,  
  follows the setup guide, and confirms readiness.}
      \label{fig:interface-setup}
  \end{figure}
  \end{center}

 Figures~\ref{fig:interface-setup} and~\ref{fig:interface-task}
  show the task page used in our user study. Each session        
  consists of five tasks, all sharing the same page layout; only
  the task prompt, reference files, and assigned agent differ. At
   the top, the page displays the task number and the assigned   
  agent. A setup panel asks whether the participant has used the 
  agent before, with three options: ``Never used,'' ``Used a few 
  times,'' and ``Use regularly.'' A collapsible guide walks      
  through installation. The participant checks a box to confirm  
  that the agent is ready. If setup fails, the participant can   
  contact the research team and skip the task after uploading    
  screenshots of the issue.

  Below the setup panel, the full task prompt is displayed.      
  Reference files are listed by name and can be downloaded       
  individually or as a ZIP archive. Expected deliverable
  filenames are listed so that participants know what to submit. 
  At the bottom of the page, participants upload their completed 
  files and provide a collaboration log showing how they worked  
  with the agent. For Manus, participants paste a shared session 
  link; for terminal-based agents like Claude Code, participants 
  upload log files or screenshots of the interaction.

\vspace*{\fill}
\vspace*{\fill}

 \newpage
\vspace*{\fill}
\begin{center}
    \captionsetup{type=figure}
    \includegraphics[width=1\textwidth]{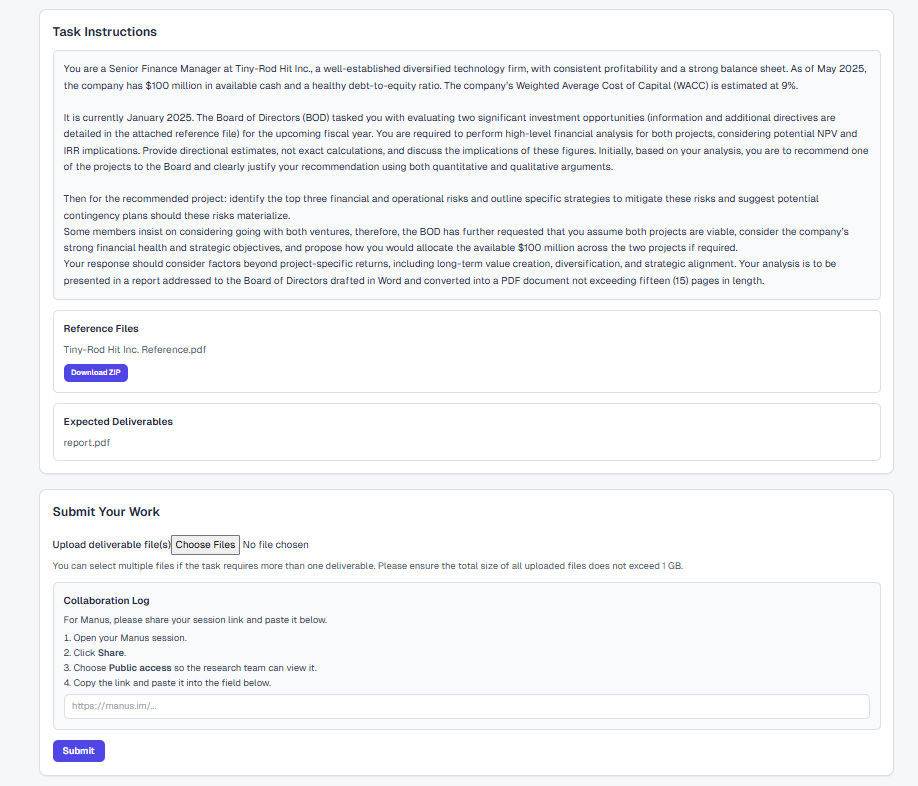}
    \captionof{figure}{\textbf{Task instruction and submission area.} The participant reads the prompt, downloads reference files, uploads deliverables, and provides a collaboration log.}
    \label{fig:interface-task}
\end{center}
\vspace*{\fill}
\newpage

\section{Prompts Used in the Automated Grader}                                                                              
  \label{appendix:prompts}                                                   
  The automated grader in \name{} works in two stages. First, a coding agent reads the task prompt and reference files to produce a structured rubric. Second, two LLM judges from different model families independently score the submitted deliverables against that rubric, and the final score is their average. 
  \refsec{sec:autograder} describes the pipeline in detail; below 
  we include the full prompts used in each stage.         

\vspace{2em}
  \begin{agentbox}{Autogenerate Rubrics Prompt}
\small

You are analyzing a task folder to generate a grading rubric for AI agent evaluation.

\vspace{0.5em}
\noindent \textbf{Task ID:} \texttt{\{task\_id\}}

\noindent \textbf{Task directory:} \texttt{\{task\_dir\}}

\noindent \textbf{Reference files available} (DO NOT read or reference anything in deliverables/): \texttt{\{file\_list\}}

\vspace{0.75em}
\noindent \textbf{INSTRUCTIONS}
\begin{enumerate}
    \item Read \texttt{prompt.txt} to understand the full task description and requirements.
    \item Read all other reference files listed above. These are the source data and templates the agent uses.
    \item Extract specific quantitative values, formulas, thresholds, counts, and required outputs from these files.
    \item Generate a \texttt{rubric.json} that:
    \begin{itemize}
        \item has \texttt{task\_id = "\{task\_id\}"}
        \item has \texttt{total\_points = 100}
        \item is split into exactly 4 scoring categories:
        \begin{itemize}
            \item \texttt{"Correctness \& Accuracy"} (40 pts)
            \item \texttt{"Deliverable Completeness"} (20 pts)
            \item \texttt{"Task Requirements"} (25 pts)
            \item \texttt{"Technical Quality"} (15 pts)
        \end{itemize}
        \item has each category as an object with keys: \texttt{name}, \texttt{max\_points}, \texttt{criteria} (list of criterion objects)
        \item has each criterion object include: \texttt{description}, \texttt{points}, \texttt{expected\_value}, \texttt{evaluation\_method}
        \item has \texttt{expected\_answers} contain \texttt{specific\_values} with exact expected outputs extracted from the reference files
        \item has \texttt{critical\_requirements} list non-negotiable pass/fail requirements
        \item has \texttt{reference\_files\_analyzed} list every file you read
        \item has \texttt{evaluation\_notes} give overall guidance for the human evaluator
    \end{itemize}
    \item Be as specific as possible. If the task requires calculating a sample size, cost, threshold, count, or any numeric value, state the exact expected answer in \texttt{expected\_value}.
    \item Output only the raw JSON object, with no markdown, no explanation, and no code fences.
\end{enumerate}
\vspace{0.75em}
\noindent \textbf{FINAL INSTRUCTION}

Now read the files and generate the \texttt{rubric.json} content.

\end{agentbox}

 \captionof{figure}{\textbf{Prompt for rubric generation.} The coding agent reads the task prompt and reference files, extracts       
  expected values from the source materials, and outputs a JSON rubric with four weighted scoring categories. The agent does
  not see any submitted deliverables at this stage.}                                  
\label{prompt:rubric_generation}

\clearpage
  \begin{agentbox}{Autograder Prompt}
\small

You are an expert evaluator scoring an AI agent's task completion against a rubric.

\vspace{0.5em}
\noindent \textbf{TASK DESCRIPTION:} \texttt{\{task\_prompt\}}

\noindent \textbf{Deliverable files submitted by the agent:} \texttt{\{file\_list\}}

\noindent \textbf{RUBRIC:} \texttt{\{rubric\_str\}}

\vspace{0.75em}
\noindent \textbf{INSTRUCTIONS}
\begin{enumerate}
    \item Read every file in the deliverables directory listed above.
    \item Use the task description above to understand what the agent was supposed to accomplish.
    \item For each rubric category and criterion, carefully evaluate whether the agent's output meets the expected value. Be specific about what you found versus what was expected.
    \item Assign a numeric score for each criterion, from 0 up to that criterion's full point value. Partial credit is allowed where the work is partially correct.
    \item Sum the criterion scores within each category.
    \item Sum all category scores for the total score.
    \item Be quantitatively precise. If a rubric criterion asks for a specific numeric value (for example, sample size = 385 or cost = \$1,234.56), check for that exact value and deduct points proportionally if it is incorrect.
\end{enumerate}

\vspace{0.75em}
\noindent \textbf{OUTPUT FORMAT}

Return only a valid JSON object with this exact structure, with no markdown and no explanation:

\begin{quote}
\ttfamily
\{
\par "score": <total numeric score, float>,
\par "max\_score": 100,
\par "percentage": <score / 100 * 100, float>,
\par "summary": "<2--4 sentence overall assessment of the agent's submission>",
\par "category\_scores": \{
\par \hspace*{1em}"<Category Name>": \{
\par \hspace*{2em}"earned": <float>,
\par \hspace*{2em}"max": <int>,
\par \hspace*{2em}"notes": "<specific observations about what was correct/incorrect in this category>"
\par \hspace*{1em}\}
\par \},
\par "deliverables\_found": [<list of filenames you were able to read>],
\par "criterion\_scores": [
\par \hspace*{1em}\{
\par \hspace*{2em}"category": "<category name>",
\par \hspace*{2em}"description": "<criterion description>",
\par \hspace*{2em}"earned": <float>,
\par \hspace*{2em}"max": <int>,
\par \hspace*{2em}"reason": "<brief explanation>"
\par \hspace*{1em}\}
\par ]
\par \}
\end{quote}

\end{agentbox}

\captionof{figure}{\textbf{Prompt for autograder scoring.} Each LLM judge receives the task description, the generated rubric, and
  the submitted deliverables. The judge scores each rubric criterion with partial credit and returns a structured JSON score  
  breakdown.}                                                               
\label{prompt:scratchpad_update}

\clearpage
\section{Prompt Used For Autonomous Agents}
\begin{agentbox}{Prompt for Testing Autonomous Agents}
\small

You are an AI assistant helping to complete a real-world task. Follow these guidelines:

\begin{itemize}
    \item {Never use special characters like ‑ (U+2011), use - (U+002D) instead}
    \item {Avoid emojis and nonstandard characters}
    \item {For PDFs, always use LibreOffice to create them}
    \item {Use cross-platform fonts like Noto Sans/Noto Serif}
    \item {Deliverable text should be concise (4 sentences max)}
    \item {Always check formatting before submitting}
    \item {Add CONFIDENCE[XX] on the last line (0-100)}
    
\end{itemize}

CRITICAL FILE LOCATION REQUIREMENTS:

\begin{itemize}
    \item {You are working in a task directory. Save ALL files directly here (in the current directory).}
    \item {Use ONLY relative filenames: 'output.pdf', 'report.docx', 'analysis.xlsx' - NO paths!}
    \item {NEVER create directories like 'deliverable\_files' or navigate to other directories}
    \item {NEVER use paths like '../deliverable\_files/' or '/deliverable\_files/' or any absolute paths}
    \item {NEVER use paths like '../deliverable\_files/' or '/deliverable\_files/' or any absolute paths}
    \item {When using LibreOffice or any tool, specify output file as just the filename: 'output.pdf' not './deliverable\_files/output.pdf'}
    \item {All files you create must appear in the same directory as prompt.txt}
    \item {Do NOT use cd, mkdir, or navigate away from the current working directory}
\end{itemize}

\{Task Prompt\}
\end{agentbox}

\captionof{figure}{\textbf{Autonomous agent prompt.} System prompt used to evaluate terminal-based agents in headless mode, engineered by AI researchers.}
\label{prompt:autonomous_agent}

\section{Deriving Bayesian Skill Rating System}
In \refsec{sec:ranking}, we introduce how \name employs a Bayesian Skill Rating System to jointly estimate human and agent skill from team outcomes. Here we provide additional derivations to assist the understanding.

\subsection{Deriving the Update from One Data Point}
\label{appendix:deriving_update}

For each teamwork outcome observation $(A, H, y)$ collected through the \name infrastructure, consider the two-dimensional latent state
$\theta = [s_A, s_H]^\top$, with prior $\theta \sim \mathcal{N}(m,\Sigma)$. Under the additive latent-skill decomposition, the observed teamwork outcome $y = x^\top \theta + \epsilon$, where $x = [1, 1]^\top$ and $\epsilon \sim \mathcal{N}(0,\beta^2)$. The goal is to update the estimation of $m$ and $\Sigma$ from this data point.

\paragraph{Likelihood.}
Conditioned on $\theta$, the observation follows
\begin{equation}
    p(y \mid \theta) = \mathcal{N}(y \mid x^\top \theta, \beta^2).
\end{equation}

Equivalently, in exponential form,
\begin{equation}
    p(y \mid \theta)
\propto
\exp\left(
-\frac{1}{2\beta^2}(y - x^\top \theta)^2
\right).
\end{equation}

\paragraph{Posterior via Bayes' rule.}
The posterior is given by
\begin{equation}
p(\theta \mid y) \propto p(y \mid \theta)\, p(\theta),    
\end{equation}

where the prior is
\begin{equation}
p(\theta)
\propto
\exp\left(
-\frac{1}{2}(\theta - m)^\top \Sigma^{-1} (\theta - m)
\right).
\end{equation}

Expanding both terms, we obtain
\begin{align}
\log p(\theta \mid y)
&= -\frac{1}{2}(\theta - m)^\top \Sigma^{-1} (\theta - m)
   -\frac{1}{2\beta^2}(y - x^\top \theta)^2 + \text{const}\\
&= -\frac{1}{2}
\theta^\top \left(
\Sigma^{-1} + \frac{1}{\beta^2} x x^\top
\right)\theta + \left(
\Sigma^{-1} m + \frac{y}{\beta^2} x
\right)^\top \theta
+ \text{const}.
\end{align}

This corresponds to a Gaussian posterior $p(\theta \mid y) = \mathcal{N}(m', \Sigma')$, with natural parameters
\begin{equation}
\Sigma'^{-1}
= \Sigma^{-1} + \frac{1}{\beta^2} x x^\top,
\qquad
\Sigma'^{-1} m'
= \Sigma^{-1} m + \frac{y}{\beta^2} x.
\label{eq:posterior}
\end{equation}

\paragraph{Recovering the covariance.}
Applying the Woodbury matrix identity,
\begin{equation}
    \Sigma'
=
\Sigma - \Sigma x (x^\top \Sigma x + \beta^2)^{-1} x^\top \Sigma.
\end{equation}

\paragraph{Recovering the mean.}
Multiplying both sides of $\Sigma'^{-1} m'$ by $\Sigma'$ and simplifying yields
\begin{equation}
m' = m + K (y - x^\top m), \quad \text{where}\quad K = \frac{\Sigma x}{x^\top \Sigma x + \beta^2}.
\end{equation}

\paragraph{Final update equations.}
Defining the residual $r = y - x^\top m$, the posterior update takes the Kalman form~\citep{Kalman1960}:
\[
r = y - x^\top m,\qquad
K = \frac{\Sigma x}{x^\top \Sigma x + \beta^2},
\]
\[
m' = m + Kr,\qquad
\Sigma' = \Sigma - K x^\top \Sigma.
\]

\subsection{Scalable Inference Algorithm}
\label{appendix:estimating_var}
While the update in Equation (\ref{eq:posteria_update}) is intuitive, it is not the most efficient representation at scale because explicitly maintaining a dense covariance matrix becomes prohibitive as the number of entities grows. For scalable inference, we instead maintain the global posterior in information form. Let $\theta \in \mathbb{R}^n$ stack all latent skills across agents and humans. For each teamwork outcome observation $(A, H, y)$, define a sparse design vector $x \in \mathbb{R}^n$ such that $x_A = 1$, $x_H = 1$, and all other entries are zero. The observation model becomes
\begin{equation}
y = x^\top \theta + \epsilon, \qquad \epsilon \sim \mathcal{N}(0,\beta^2).
\end{equation}
In the information form, 
\begin{equation}
p(\theta) \propto \exp\!\left(-\frac12 \theta^\top \Lambda \theta + \eta^\top \theta\right),
\end{equation}
where $\Lambda=\Sigma^{-1}, \eta=\Sigma^{-1}m$. Writing $w = 1/\beta^2$, the update in Equation (\ref{eq:posteria_update}) is equivalent to the following (see Appendix~\ref{appendix:deriving_update} for derivation):
\begin{equation}
\Lambda \leftarrow \Lambda + w\, x x^\top,\qquad
\eta \leftarrow \eta + w\, y\, x.
\label{eq:information_form_update}
\end{equation}
Because $x$ has only two nonzero entries, this update is extremely sparse, touching only four entries in $\Lambda$ and two entries in $\eta$.
Priors in Equation (\ref{eq:prior}) are injected exactly once per entity by adding the corresponding Gaussian natural parameters:
\begin{equation}
    \Lambda_{ii} \mathrel{+}= 1/\sigma_0^2,\qquad
\eta_i \mathrel{+}= \mu_0/\sigma_0^2.
\end{equation}

Once all observations have been accumulated, the posterior mean vector, which serves as the primary estimate of latent skill, is obtained by solving the sparse linear system $\Lambda \mu = \eta$. Uncertainty for entity $i$ is determined by the marginal posterior variance,
\[
\sigma_i^2 = (\Lambda^{-1})_{ii},
\qquad
\sigma_i = \sqrt{(\Lambda^{-1})_{ii}}.
\]
In practice, we initialize all entities with $\mu_0=0$ and $\sigma_0=1$, and use Hutchinson's stochastic diagonal estimator~\citep{Hutchinson1990,Bekas2007} to estimate $(\Lambda^{-1})_{ii}$ efficiently.

\section{Comparing \name Rating System with Averaging Agent Performance}
\label{appendix:rating_method_comparison}

We compare \name ratings against a naive baseline that ranks agents by mean task score. \reftab{table:ranking_method_comparison} shows the results. While the resulting agent ranking order matches that of \name, the high variance ($\sigma > 19$ for all agents) makes it difficult to draw statistically meaningful conclusions due to high inter-human variability. This high variance is the core motivation for jointly modeling human skill. Furthermore, the naive baseline offers no insight into the human side, providing no information on human workers' skill in collaborating with AI agents.

\begin{table}[h]
\centering
\small
\begin{tabular}{lcc!{\vrule width \lightrulewidth}cc} 
\toprule
\multicolumn{1}{c}{\multirow{2}{*}{\textbf{Agent }}} & \multicolumn{2}{c!{\vrule width \lightrulewidth}}{\textbf{Rating by Taking the Average }} & \multicolumn{2}{c}{\textbf{Rating by Using CollabSkill Framework }}  \\
\multicolumn{1}{c}{}                                 & \textbf{Score} & $\sigma$                                                                 & \textbf{Score} & $\sigma$                                            \\ 
\midrule
Claude Cowork                                        & 83.839         & 19.456                                                                   & 76.738         & 0.156                                               \\
Claude Code                                          & 81.176         & 21.994                                                                   & 74.778         & 0.157                                               \\
Codex                                                & 77.480         & 25.286                                                                   & 71.267         & 0.155                                               \\
Manus                                                & 76.417         & 24.887                                                                   & 71.204         & 0.147                                               \\
Gemini CLI                                           & 74.822         & 22.528                                                                   & 69.555         & 0.151                                               \\
\bottomrule
\end{tabular}

\caption{Comparing CollabSkill ratings against ratings by taking the average of agent performance.}
\label{table:ranking_method_comparison}
\end{table}

\section{Survey Details}
\name evaluates human-agent collaboration with real human workers. To better understand the human factors, we collect data on participants' AI literacy and attitude through pre/post-task surveys. Our survey is structured as follows:

\paragraph{Demographics and Prior Exposure}
\begin{enumerate}[label=\textbf{D\arabic*.}, leftmargin=3.5em]
    \item \label{item:D1} \textbf{LLM Familiarity:} How familiar are you with large language model products (e.g., ChatGPT, Claude, Google Gemini, etc.)?
    \begin{itemize}
        \item I use them regularly.
        \item I have some experience using them.
        \item I have heard of them but don't know much about their functionalities.
        \item No, I've never heard of them.
    \end{itemize}

    \item \label{item:D2} \textbf{Professional Use:} Have you used large language models in your work-related activities?
    \begin{itemize}
        \item Yes, I use them every day in my work.
        \item Yes, I use them every week in my work.
        \item Yes, I have used them occasionally for specific tasks.
        \item No, I have not used them for any work-related activities.
        \item No, I've never heard of them.
    \end{itemize}

    \item \label{item:D3} \textbf{Experience:} How many years of experience do you have in [participant's occupation]? \textit{(Numeric input)}
\end{enumerate}

\paragraph{Attitudinal Readiness}
\label{app:attitude_survey}
To measure shifts in participants' perceptions of AI agents, we administered the following items both before and after the collaborative tasks. Post-task items were prefixed with: \textit{``After collaborating with agents on these tasks...''}

\begin{table}[t]
\centering
\small
\begin{tabularx}{\textwidth}{llX}
\toprule
\textbf{Level} & \textbf{Type} & \textbf{Definition} \\ \midrule
\textbf{H1} & Autonomous & AI agent handles the task entirely on its own. \\
\textbf{H2} & Minimal Input & AI agent requires minimal human input for optimal performance. \\
\textbf{H3} & Partnership & AI agent and human form an equal partnership, outperforming either alone. \\
\textbf{H4} & Human-Led & AI agent requires human input to successfully complete the task. \\
\textbf{H5} & Full Involvement & AI agent cannot function without continuous human involvement. \\ \bottomrule
\end{tabularx}
\label{tab:has_scale}

\caption{The Human Agency Scale (HAS)~\citep{shao2025future} reference provided to participants in pre/post-task surveys.}
\label{table:has}
\end{table}

\begin{enumerate}[label=\textbf{A\arabic*.}, leftmargin=3.5em]
    \item \label{item:A1} \textbf{Agent Familiarity:} How familiar are you with AI agents (systems designed to autonomously complete tasks on your behalf)? \textit{(7-point Likert: 1 = Never heard of them, 7 = Use them regularly)}

     \item \label{item:A2} \textbf{Trust in Autonomy:} How much do you trust AI agents to execute tasks accurately without human oversight? \textit{(7-point Likert: 1 = No trust at all, 7 = Complete trust)}
    
    \item \label{item:A3} \textbf{Comfort with Delegation:} How comfortable are you delegating high-stakes professional tasks to an AI agent? \textit{(7-point Likert: 1 = Very uncomfortable, 7 = Very comfortable)}
    
    \item \label{item:A4} \textbf{Perceived Capability:} Reflecting on your typical work tasks, where on the HAS scale do you believe AI agents can generally operate for your job? \textit{(Choice: H1--H5, see \reftab{table:has} for reference)}
    
    \item \label{item:A5} \textbf{Desired Autonomy:} Where on the HAS scale do you prefer AI agents to operate within your professional workflow? \textit{(Choice: H1--H5, see \reftab{table:has} for reference)}
    
    \item \label{item:A6} \textbf{Role Meaningfulness:} To what extent are you concerned that AI integration will reduce your role below what you consider meaningful work? \textit{(7-point Likert: 1 = Not at all, 7 = Extremely worried)}

    \item \label{item:A7} \textbf{Affective Orientation:} Which of the following best captures your current sentiment toward AI agents? \textit{(Options: Excited, Curious, Neutral, Skeptical, Anxious)}
\end{enumerate}

\section{Validating Automated Grader Quality}
\label{appendix:autograder_validation}
\subsection{Rubric Categories}
To evaluate open-ended deliverables, we partition evaluation criteria into four weighted top-level categories. This multi-dimensional approach is grounded in established frameworks for automated and agentic evaluation. \textit{Correctness \& Accuracy} assesses factual integrity and technical soundness, treating factual accuracy as the primary axis of evaluation~\citep{liang2022helm, anghel2025pearl}. \textit{Deliverable Completeness} measures whether the agent successfully produced all requested components, which serves as a fundamental dimension of task utility~\citep{arabzadeh2024agenteval}. \textit{Task Requirements} evaluates adherence to explicit instructions and negative constraints, ensuring the agent operates within defined operational guardrails~\citep{liang2022helm}. Finally, \textit{Technical Quality} captures organization, clarity, and professionalism, drawing on established metrics for explanatory usefulness and clarity~\citep{anghel2025pearl, arabzadeh2024agenteval}. We weight correctness highest, as the economic utility of occupational agents is fundamentally predicated on factual accuracy over stylistic alignment. \reffig{fig:example_rubric} shows an example rubric generated from the automated grader.

\subsection{Manual Rubric Audit}
For realistic occupational tasks, manual grading is hard to scale: GDPval reports that grading each comparison took human experts an hour on average~\citep{patwardhan2025gdpval}, and Remote Labor Index reports human evaluators spending around 30 minutes per task for pairwise preference judgments~\citep{mazeika2025remote}. While a more scalable approach is to use LLM-as-a-judge provided with fine-grained rubrics, many task sources do not come with rubrics or only cover correctness, and manual rubric construction remains expensive, motivating our reference-free approach where we employ an agent to generate rubrics automatically. We validate rubric quality by sampling 3 tasks per sector from GDPVal (27 tasks total) and manually comparing our auto-generated rubrics against the official hand-authored GDPVal rubrics\footnote{\url{https://huggingface.co/datasets/openai/gdpval}}, which were released after the original GDPVal paper~\citep{patwardhan2025gdpval}. The generated rubrics achieve a recall of 82.1\% against the official criteria (998 of 1,216 official rubric items are covered) and a precision of 92.2\% (297 of 322 of our items are grounded in official criteria), with authors judging all 322 of the generated rubric items as important for evaluating task quality.%

\subsection{Correlation with Grading with Hand-authored Rubrics}
We further validate our grading pipeline by comparing the final scores yielded from our automated grader and the scores yielded from GDPval official rubrics on 135 randomly sampled (task, deliverable) pairs, with authors manually reviewing each score to confirm evaluation outcomes The two score series yielded a Pearson correlation of $r=0.72$ (excluding three outlier pairs based on author review; $r = 0.66$ over all 135 pairs), indicating strong agreement and supporting the use of auto-generated rubrics in the automated grading pipeline.

\begin{examplebox}{Example Rubric - Istanbul Trip Itinerary}
\small

\textbf{Task ID:} \texttt{0e4fe8cd-16d0-4f41-8247-6385b4762582}

\vspace{0.5em}
\noindent \textbf{Total Points:} 100

\vspace{0.75em}
\noindent \textbf{SCORING CATEGORIES}

\vspace{0.5em}
\noindent \textbf{1. Correctness \& Accuracy} \hfill \textit{40 pts}
\begin{itemize}
    \item \textbf{Itinerary dates are correct (June 1--4)} \hfill 5 pts \\
    \textit{Expected:} Tabs labeled June 1st, June 2nd, June 3rd, June 4th. \\
    \textit{Method:} Manual inspection of tab names and dates.

    \item \textbf{All specified times are correct} \hfill 10 pts \\
    \textit{Expected:}
    \begin{itemize}
        \item Day 1: 8am Pickup, 9am Wheels Up
        \item Day 2: 3:00am Wheels Down, 4:30am Drop-off, 9:00am Breakfast, 11:00am Tour, 2:00pm Lunch, 5:00pm Return/Meeting, 7:00pm Tuxedo Drop-off, 8:30pm Pickup, 9:00pm Dinner, 10:30pm Departure
        \item Day 3: 9:00am Breakfast, 11:00am Hair/Makeup, 1:00pm Tuxedo Fitting, 2:00pm Lunch, 4:00pm SUV to Wedding, 10:30pm Leave Wedding
        \item Day 4: 8:00am Pickup, 9:00am Wheels Up, 11:00am Landing, 11:00am Pickup, 11:30am Drop-off
    \end{itemize}
    \textit{Method:} Compare times listed against expected values.

    \item \textbf{All locations are correctly named} \hfill 10 pts \\
    \textit{Expected:} Airports: JVY, ISL. Hotel: Four Seasons Bosphorus. Restaurants: Yali, Hidden Garden, Garden 1897. Venue: Adile Sultan Palace. \\
    \textit{Method:} Verify all location names against the prompt.

    \item \textbf{All names are correctly spelled} \hfill 5 pts \\
    \textit{Expected:} Oguz (Tour Guide), Maserto (Tuxedo), Samira Lowell (Hair/Makeup), Bespoke Tuxedo. \\
    \textit{Method:} Check spelling of all specified names.

    \item \textbf{Flight calculations and time zone details are accurate} \hfill 10 pts \\
    \textit{Expected:} Outbound: 10-hour flight, 8-hour time difference forward. Return: 10-hour flight, gaining 8 hours. \\
    \textit{Method:} Confirm flight duration and time zone details in travel entries.
\end{itemize}

\vspace{0.5em}
\noindent \textbf{2. Deliverable Completeness} \hfill \textit{20 pts}
\begin{itemize}
    \item \textbf{Workbook contains exactly four tabs} \hfill 10 pts \\
    \textit{Expected:} 4 tabs, one per day. \\
    \textit{Method:} Count the number of tabs in the submitted Excel file.

    \item \textbf{All required hyperlinks are included} \hfill 10 pts \\
    \textit{Expected:} Links for: Oguz (tour guide), Hidden Garden, Four Seasons Bosphorus GM, Maserto, Garden 1897, Samira Lowell, and Adile Sultan Palace. \\
    \textit{Method:} Check for the presence of a hyperlink for each specified item.
\end{itemize}

\vspace{0.5em}
\noindent \textbf{3. Task Requirements} \hfill \textit{25 pts}
\begin{itemize}
    \item \textbf{Itinerary delivered as a multi-tab Excel document} \hfill 10 pts \\
    \textit{Expected:} File format \texttt{.xlsx} or equivalent. \\
    \textit{Method:} Verify the file type of the deliverable.

    \item \textbf{Itinerary covers complete journey door-to-door} \hfill 5 pts \\
    \textit{Expected:} Starts with pickup at the main house (Day 1) and ends with drop-off at the main house (Day 4). \\
    \textit{Method:} Check the start and end points of the itinerary.

    \item \textbf{Separate actions listed as distinct line items} \hfill 5 pts \\
    \textit{Expected:} E.g., ``8:30 p.m. picked up from hotel lobby'' and ``9:00 p.m. reservation at Garden 1897'' are two separate entries. \\
    \textit{Method:} Inspect itinerary to ensure events are broken into individual line items.

    \item \textbf{Formatting is readable and all links are clickable} \hfill 5 pts \\
    \textit{Expected:} Clear, professional layout; hyperlinks are active and lead to relevant webpages. \\
    \textit{Method:} Manual review and hyperlink testing.
\end{itemize}

\vspace{0.5em}
\noindent \textbf{4. Technical Quality} \hfill \textit{15 pts}
\begin{itemize}
    \item \textbf{File is a valid, openable Excel workbook} \hfill 5 pts \\
    \textit{Expected:} Opens without errors in standard spreadsheet software. \\
    \textit{Method:} Attempt to open the file.

    \item \textbf{Workbook is well-organized with clear columns} \hfill 5 pts \\
    \textit{Expected:} Columns for Time, Action/Event, Details/Links clearly labeled and consistently used. \\
    \textit{Method:} Subjective review of layout and organization.

    \item \textbf{No placeholder, broken, or incorrect links} \hfill 5 pts \\
    \textit{Expected:} All hyperlinks resolve to valid, relevant websites; high-quality alternatives provided for undiscoverable entities. \\
    \textit{Method:} Click-through on all hyperlinks to verify validity and relevance.
\end{itemize}

\vspace{0.75em}
\noindent \textbf{EXPECTED SPECIFIC VALUES}
\begin{center}
\begin{tabular}{ll}
\hline
\textbf{Item} & \textbf{Expected Value} \\
\hline
Day 1 Start Time & 8:00 AM \\
Day 1 Departure Airport & JVY \\
Day 2 Arrival Airport & ISL \\
Hotel Name & Four Seasons Bosphorus \\
Tour Guide Name & Oguz \\
Wedding Venue & Adile Sultan Palace \\
Number of Excel Tabs & 4 \\
\hline
\end{tabular}
\end{center}

\vspace{0.75em}
\noindent \textbf{CRITICAL REQUIREMENTS}
\begin{enumerate}
    \item The deliverable must be a multi-tab Excel file (\texttt{.xlsx} or equivalent).
    \item The itinerary must span the four specified days (June 1--June 4).
    \item The itinerary must include travel logistics from the principal's home and back.
\end{enumerate}

\vspace{0.75em}
\noindent \textbf{EVALUATION NOTES}

The core of this task is attention to detail and professional presentation. The evaluator should check every time, location, and name against the prompt. The quality of research for hyperlinks is also important; links should be for high-end, relevant services in Istanbul. If a specified entity such as \textit{Maserto} cannot be found, the agent should have noted this and provided a comparable high-quality alternative.

\end{examplebox}
\captionof{figure}{\textbf{Example rubric for the Istanbul trip itinerary task.} The rubric evaluates correctness of times, locations, and names; completeness of tabs and hyperlinks; adherence to formatting requirements; and overall technical quality of the Excel deliverable.%
}
\label{fig:example_rubric}

\section{Analysis of Human-Agent Collaboration Trajectories}
\label{sec:trajectory_analysis}
To characterize how participants interact with AI agents during collaboration sessions, we conducted a systematic analysis of all available interaction logs ($N=386$). We code each session's trajectory with \texttt{gemini-3-pro} using the 11 behavioral indicators from the Anthropic AI Fluency Index, which captures effective human–AI collaboration practices such as iterating on outputs, clarifying goals, specifying formats, defining audience, and verifying AI-generated claims (see \reffig{prompt:trajectory_analysis} for the full prompt).

We examined the relationship between each AI fluency behavior and the scalar outcome of the session (mean=78.6, N=386) using point-biserial correlations. Figure~\ref{Fig:ai_fluency_correlation} presents the results. Six of the eleven behaviors were significantly correlated with higher session scores, including (1) defining the audience for the output ($r=0.345$, $p<0.001$), where sessions exhibiting this behavior scored 16.5 points higher on average (84.7 vs. 68.1), (2) communicating tone and style preferences ($r=0.232$, $p<0.001$; 83.9 vs. 73.2), (3) providing examples of desired output quality ($r=0.199$, $p<0.001$; 82.8 vs. 73.5), (4) setting the interaction mode ($r=0.167$, $p=0.001$; 82.6 vs. 74.9), (5) clarifying the goal before requesting help ($r=0.141$, $p=0.006$; 79.8 vs. 69.6), and (6) specifying format and structure ($r=0.121$, $p=0.017$; 79.6 vs. 71.1). Notably, the raw number of user turns was uncorrelated with session score ($r=0.046$, $p=0.371$), suggesting that more interaction does not inherently lead to better outcomes and the quality of the interaction matters more.

We also compare AI fluency behavior rates across the five agents in our study. Cowork elicits the highest mean AI fluency score (6.0 out of 11 on average). Examining individual behaviors, Cowork led all agents in eliciting the human workers to set interaction mode (60.3\% vs. 39–49\% for other agents), identify missing context (83.6\%), question reasoning (16\%), consult on approach (8.2\%), and check facts (28.8\%).

\newpage
\begin{agentbox}{Collaboration Trajectory Analysis Prompt}
\small
You are an expert at analyzing human-AI interaction patterns.
You will be given a conversation log between a user and an AI assistant.

Your task is to analyze whether the user demonstrates the following effective AI collaboration behaviors. For each behavior, assess with a simple Yes or No based on whether there is clear evidence in the log, and provide a brief reasoning for your assessment.

---

BEHAVIORS TO ASSESS:
\begin{enumerate}
    \item {**Iterates and Refines** — Does the user follow up, push further, or build on prior responses rather than accepting the first answer?}

    \item{**Clarifies Goal Before Asking for Help** — Does the user articulate what they're trying to achieve before making requests?}

    \item{**Provides Examples of What Good Looks Like** — Does the user share reference points, models, or success criteria?}

    \item{**Specifies Format and Structure Needed** — Does the user indicate how they want the output organized (length, format, sections, etc.)?}

    \item{**Sets Interaction Mode** — Does the user define how they want the AI to behave (e.g., "be concise", "act as a critic", "ask me questions before answering")?}

    \item{**Communication Tone and Style Preferences** — Does the user express stylistic preferences (formal/casual, direct/exploratory, etc.)?}

    \item{**Identifies When AI Might Be Missing Context** — Does the user proactively add background, correct assumptions, or flag gaps in the AI's understanding?}

    \item{**Defines Audience for the Output** — Does the user specify who the output is for and tailor the request accordingly?}

    \item{**Questions When AI Reasoning Doesn't Hold Up** — Does the user push back, probe the AI's logic, or challenge weak reasoning?}

    \item{**Consults AI on Approach Before Execution** — Does the user ask the AI to help think through strategy or method before diving in?}

    \item{**Checks Facts and Claims That Matter** — Does the user verify outputs, ask for sources, or flag uncertainty about specific claims?}
\end{enumerate}

---

OUTPUT FORMAT:

Return only a valid JSON object. No explanation, no markdown, no preamble.
Each behavior has two fields: "result" ("yes" or "no") and "reasoning"
(one concise sentence explaining why).

\begin{quote}
\ttfamily
\{
\par \hspace*{1em}"iterates\_and\_refines": \{"result": "", "reasoning": ""\},
\par \hspace*{1em}"clarifies\_goal\_before\_asking": \{"result": "", "reasoning": ""\},
\par \hspace*{1em}"provides\_examples\_of\_good": \{"result": "", "reasoning": ""\},
\par \hspace*{1em}"specifies\_format\_and\_structure": \{"result": "", "reasoning": ""\},
\par \hspace*{1em}"sets\_interaction\_mode": \{"result": "", "reasoning": ""\},
\par \hspace*{1em}"communication\_tone\_and\_style": \{"result": "", "reasoning": ""\},
\par \hspace*{1em}"identifies\_missing\_context": \{"result": "", "reasoning": ""\},
\par \hspace*{1em}"defines\_audience": \{"result": "", "reasoning": ""\},
\par \hspace*{1em}"questions\_ai\_reasoning": \{"result": "", "reasoning": ""\},
\par \hspace*{1em}"consults\_ai\_on\_approach": \{"result": "", "reasoning": ""\},
\par \hspace*{1em}"checks\_facts\_and\_claims": \{"result": "", "reasoning": ""\}
\par \}
\end{quote}

---

Here is the interaction log to analyze:

\{Interaction Log\}

\end{agentbox}
\vspace{-1em}
\captionof{figure}{\textbf{Prompt for coding AI fluency behavior indicators.} We parse the human-agent collaboration log into text format and use an LLM to code behaviors from the Anthropic AI Fluency Index which identifies 11 directly observable indicators of human skill in using AI~\citep{anthropic2026aifluency}.}                                                               
\label{prompt:trajectory_analysis}

\end{document}